\documentclass[aps,prl,showpacs,twocolumn,superscriptaddress]{revtex4}
\usepackage{graphicx}
\usepackage{bm}
\usepackage{amsmath}
\usepackage{amsfonts}

\def\ra{\rangle}
\def\la{\langle}
\def\be{\begin{equation}}
\def\ee{\end{equation}}
\def\bea{\begin{eqnarray}}
\def\eea{\end{eqnarray}}

\def\ua{\uparrow}
\def\da{\downarrow}
\def\dag{^\dagger}
\def\sig{\sigma}
\def\om{\omega}

\begin{document}
\title{Quantum Dynamics of Multiferroic Helimagnets: a Schwinger-Boson Approach}

\author{Hosho Katsura}
\email{katsura@appi.t.u-tokyo.ac.jp}
\affiliation{Department of Applied Physics, The University of Tokyo,
7-3-1, Hongo, Bunkyo-ku, Tokyo 113-8656, Japan}

\author{Shigeki Onoda}
\affiliation{Condensed Matter Theory Laboratory, RIKEN, Wako, Saitama 351-0198, 
Japan}

\author{Jung Hoon Han}
\affiliation{BK21 Physics Research Division, Department of Physics, 
Sungkyunkwan University, Suwon 440-746, Korea}

\author{Naoto Nagaosa}
\affiliation{Department of Applied Physics, The University of Tokyo,
7-3-1, Hongo, Bunkyo-ku, Tokyo 113-8656, Japan}
\affiliation{Cross Correlated Materials Research Group, Frontier Research System, Riken,2-1 Hirosawa, Wako, Saitama 351-0198, Japan }

\begin{abstract}
We study the quantum dynamics/fluctuation of the cycloidal helical magnet in terms of the Schwinger boson approach. In sharp contrast to the classical fluctuation, the quantum fluctuation is collinear in nature which gives rise to the collinear spin density wave state slightly above the helical cycloidal state as the temperature is lowered. 
Physical properties such as the reduced elliptic ratio of the spiral, the neutron scattering and infrared absorption spectra are discussed from this viewpoint with the possible relevance to the quasi-one dimensional LiCu$_2$O$_2$ and LiCuVO$_4$. 
\end{abstract}

\maketitle
Frustration, competition between interactions, in magnets has been an intriguing issue in the field of classical/quantum magnetism over several decades. 
In the usual case, even with the competing exchange interactions $J_{ij}$'s, their Fourier transform  $J(q)$ has the maximum at some wavevector $q=Q$, and the classical ground state becomes the helimagnet \cite{Yoshimori}. 
This is because of the constraint of the fixed length of the classical spin, i.e., $|{\bm S}_j| =$fixed. 
In strongly frustrated quantum magnets, on the other hand, the long-range order is possibly destroyed and novel ground states without magnetic order may be realized. 
Many possibilities such as chiral spin liquid \cite{Laughlin}, spin-nematic \cite{Chandra90} and spin-Peierls/valence-bond-crystal \cite{Read_Sachdev} states are theoretically proposed. 
Another possibility is a magnetically ordered state realized by the order-by-disorder mechanism when the corresponding classical system has continuously degenerate ground states \cite{Henley}.

Recently a renewed interest has been focused on the cycloidal helimagnets from the viewpoint of {\it multiferroics}, which exhibit both magnetic and ferroelectric properties \cite{Fiebig, Tokura}.
These materials shed some new light on the frustrated magnets since the electric polarization is closely related to the vector spin chirality ${\bm S}_i \times {\bm S}_j$\cite{KNB, Dagotto, Jia,Mostovoy,Harris}, which has been the subject of intensive interests. 
Namely, it was found that the electric polarization($\bm P$) produced by the neighboring spins (${\bm S}_i$ and ${\bm S}_j$) can be written as 
\begin{equation}
{\bm P}=a {\bm e}_{ij}\times ({\bm S}_i \times {\bm S}_j),
\label{KNB}
\end{equation}
where ${\bm e}_{ij}$ denotes the unit vector connecting the sites $i$ and $j$. This relation has a physical interpretation in terms of spin current induced between noncollinear spins due to frustration \cite{KNB}.

Magnetic materials with the finite vector spin chirality include wide range of systems such as three dimensional(3D) magnets $R$MnO$_3$ ($R=$Gd, Tb, Dy) with spin $S=2$ \cite{KimuraNature, Goto, Noda, Kenzelmann}, the kagome staircase compound Ni$_3$V$_2$O$_8$ with $S=1$ \cite{Lawes_Ni3V2O8},  $S=1/2$ quantum spin chains LiCu$_2$O$_2$ \cite{Cheong_LiCu2O2,Seki}, LiCuVO$_4$ \cite{Sato_LiCuVO4} and the quasi-one-dimensional(1D) molecular helimagnet with $S=7/2$ \cite{Cinti_molecular_helimagnet}. 
Depending on the temperature, dimensionality, and the magnitude of the spin $S$, the role of the classical/quantum spin fluctuations differs and the theoretical studies on these fluctuations are needed for the consistent interpretation of the phase diagram and the physical properties of these systems. 
Especially, the low dimensionality enhances both thermal and quantum fluctuations leading to the breakdown of the conventional (classical + spin wave) picture for helimagnets. 
The possible chiral spin states without the magnetic long range ordering have been proposed theoretically for classical~\cite{Villain,MiyashitaShiba84,Sonoda} and quantum~\cite{Nersesyan,Hikihara,Furukawa} spin systems. 
However, the systematic study of the quantum fluctuation in the helimagnets including the finite temperature effect is rare, which is addressed in this paper and will be complementary to the works mentioned above. 

In this paper, we study the quantum/thermal fluctuation in the helimagnet in terms of the Schwinger Boson (SB) approach. 
The advantage of the SB method is that it can describe the length of the 
ordered moment as a soft variable. 
Namely, in the constraint on the Schwinger boson number at each site,
\begin{equation}
\sum_\sigma b^\dagger_{j \sigma} b_{j \sigma} = 2S,
\label{const_site}
\end{equation}
it can be decomposed into the condensed (classical) part and the fluctuating part. 
Therefore, the degrees of classical/quantum fluctuation and the ordered moment can be described in a unified fashion in this method \cite{Auerbach}. 
In the SB language, the paramagnetic to collinear transition is described by the density wave instability of bosons, while the collinear to helical one corresponds to the Bose-Einstein condensation(BEC) of SB. 
\begin{figure}
\includegraphics[width=7cm,clip]{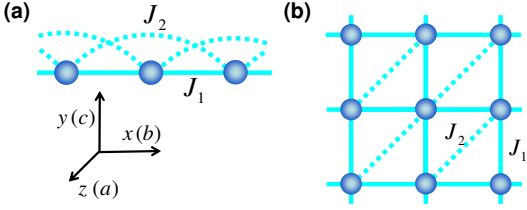}
\caption{Schematic lattice structure and exchange interactions of the effective spin model. $J_1$ is a ferromagnetic, while $J_2$ is antiferromagnetic interactions. $xyz$-coordinates and $abc$-axes are also shown.}
\label{q1d2d_lattice}
\end{figure}

\begin{figure}
\includegraphics[width=6cm,clip]{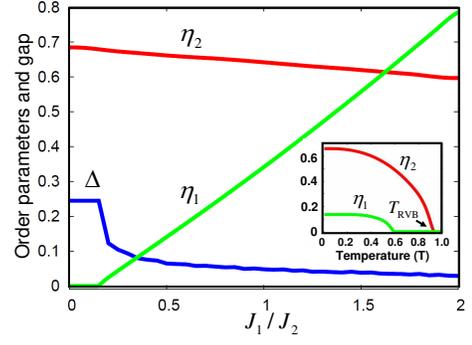}
\caption{RVB order parameters $\eta_1$ and $\eta_2$ and the gap $\Delta$ of the $S=$1/2 1D model with varying $J_1/J_2$ at zero temperature. Inset shows the temperature dependence of $\eta_1$ and $\eta_2$ at $J_1/J_2=0.5$. We use the unit $J_2=k_{\rm B}=1$.}
\label{quasi1d_orders}
\end{figure}

{\it Effective model}---We study quasi-1D and two-dimensional (2D) Heisenberg models with the exchange interactions shown in Fig. \ref{q1d2d_lattice}, where $J_1$ is ferromagnetic, while $J_2$ are antiferromagnetic, leading to the frustration. The interchain/interplane interaction $J_{\perp}$ is assumed to be sufficiently weak, and will be treated by the mean field theory.  
The spin-$S$ operators are represented by SB as 
$S^\alpha = b_{\sig}\dag(\sig^{\alpha}_{\sig\sig'}/2) b_{\sig'}$, where $\sig^\alpha$ ($\alpha=x,y,z$) are the Pauli matrices and the repeated indices are summed over. 
First, we assume that the resonating-valence-bond(RVB) correlation is dominant and neglect the other mean-field decoupling. 
This assumption is valid for the low-dimensional multiferroics~\cite{Mochizuki} such as LiCuVO$_4$ \cite{Sato_LiCuVO4}, LiCu$_2$O$_2$ \cite{Cheong_LiCu2O2} and Ni$_3$V$_2$O$_8$ \cite{Lawes_Ni3V2O8}. 
The mean-field Hamiltonians of the quasi-1D model is given by
\begin{eqnarray}
&&H^{\rm MF}_{1D} = \sum_{\bm k \sigma}(\lambda -2\eta_1 \cos k_x)b\dag_{\bm{k}\sig}b_{\bm{k}\sig} \nonumber \\
&+&\sum_{\bm k} 2[\eta_2 \sin (2k_x)+\eta_{\perp}(\sin k_y +\sin k_z )]b\dag_{\bm{k}\ua}b\dag_{-\bm{k}\da}+{\rm h.c.} \nonumber \\
&+&2 {\cal N}(\eta^2_1/J_1 + \eta^2_2/J_2 + 2 \eta^2_{\perp}/J_{\perp}-S\lambda),
\end{eqnarray}
where ${\cal N}$ is the total number of sites and $b_{k\sig}$ is the Fourier transform
defined by $ b_{j\sig}=\sum_{\bm k} e^{-i\bm{k}\cdot \bm{R}_j}b_{\bm{k}\sig}/\sqrt{\cal N}$. 
In $H^{\rm MF}_{1D}$, $\lambda$ denotes the chemical potential for the bosons and the order parameters $\eta_1$, $\eta_2$ and $\eta_{\perp}$ are $J_1\la b\dag_{i\sig}b_{i+{\hat x},\sig} \ra/2$, 
$J_2\la b_{i\nu}\epsilon_{\nu \sig}b_{i+2{\hat x},\sig} \ra/(2i)$ and 
$J_{\perp}\la b_{i\nu}\epsilon_{\nu \sig}b_{i+{\hat e},\sig} \ra/(2i)$ ($\hat e=$ $\hat y, \hat z$), respectively, with $\epsilon_{\da \ua}=-\epsilon_{\ua \da}=1$. 
RVB order parameters are assumed to be real and spatially uniform.   
In a parallel way, we can derive the quasi-2D mean-field Hamiltonian $H^{\rm MF}_{2D}$. 
The Hamiltonian $H^{\rm MF}_{1D}$ can be diagonalized by the 
Bogoliubov transformation as
$
H^{\rm MF}_{1D}=\sum_{\bm{k}\sig}\om(\bm{k}) (\gamma 
\dag_{\bm{k}\sig}\gamma_{\bm{k}\sig}+1/2)
-2{\cal N}\lambda (S+1/2)+{\rm const.},
$
with the dispersion relation
$\om({\bm k})^2=(\lambda-2\eta_1\cos k_x)^2-(2\eta_2 \sin(2k_x)+2\eta_{\perp}(\sin k_y+\sin k_z))^2$. 
The transformation between $\gamma_{\bm{k}\sig}$ and $b_{\bm{k}\sig}$ is given by
\begin{equation}
\bigg( \begin{array}{c} b_{\bm{k}\ua} \\ b_{-\bm{k}\da}\dag \end{array} \bigg)
=
\bigg( \begin{array}{cc} \cosh \theta_{\bm k} & \sinh\theta_{\bm k} 
\\ \sinh\theta_{\bm k} & \cosh\theta_{\bm k}\end{array}\bigg)
\bigg( \begin{array}{c} \gamma_{\bm{k}\ua} \\ \gamma_{-\bm{k}\da}\dag 
\end{array} \bigg),
\label{bogol}
\end{equation}
with 
$\tanh 2\theta_{\bm k}=-(2\eta_2 \sin k_x+2\eta_{\perp}(\sin k_y +\sin k_z))/(\lambda-2\eta_1 \cos k_x)$. 
The chemical potential $\lambda$ is determined by the 
condition (\ref{const_site}).
$\eta$'s are obtained by minimizing the mean-field free energy $F^{\rm MF}$. 
Figure \ref{quasi1d_orders} shows the numerically obtained $\eta_1$, $\eta_2$ and the gap $\Delta(T)=\omega({\bm Q}/2)$ of the 1D spin-1/2 model as a function of $J_1/J_2$ \cite{collapse}. 
The transition temperature of $\eta_2$ is analytically given by $T_{\rm RVB}=J_2(S+1/2)/\ln(1+1/S)$.
We have also numerically studied the 2D model at finite temperature and obtained similar results. 
From $\eta$'s, we can estimate the minima of the dispersion $\om(\bm{k})$ as  $\pm\bm{Q}/2=\pm(Q/2, \pi/2, \pi/2)$. 
$Q$ is determined to satisfy $(\lambda-2\eta_1 \cos(Q/2)) \eta_1 \sin(Q/2)=4(\eta_2\sin Q+2\eta_{\perp})\eta_2\cos Q$. 

\begin{figure}
\includegraphics[width=8cm,clip]{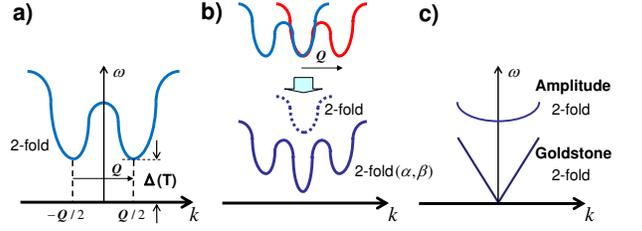}
\caption{a) Schematic energy dispersion of $\gamma$-particles. Minima are at ${\bm k}=\pm 
\bm{Q}/2$. b) The reorganization of the SB due to the collinear magnetic order.
The origin of the momentum ${\bm k}$ is shifted by $\pm{\bm Q}/2$. c) Goldstone and amplitude modes associated with the BEC in the helical phase.
}
\label{disp}
\end{figure}

To describe the low-energy physics of the model, it is useful to construct an effective continuum model. First, we suppose that $\eta$'s are non-zero. 
Next we expand the dispersion around the minima up to quadratic order in 
$\bm{k}\pm\bm{Q}/2$.  The effective dispersion relations of $\gamma$-particles are those of massive relativistic bosons and explicitly given by
$\Omega({\bm k})=\sqrt{\Delta(T)^2+c^2_{\parallel} |{\bm k}_{\parallel}|^2+c^2_{\perp} |{\bm k}_{\perp}|^2}$, where ${\bm k}_{\parallel}$ is the vector along (within) the chain (plane) while ${\bm k}_{\perp}$ is that perpendicular to the chain (plane).  
The spin wave velocities $c_{\parallel}$ and $c_{\perp}$  can be written in terms of $\eta$'s, in principle. Now the effective Hamiltonian of our system is
\begin{equation}
H^{\rm eff}=\sum_{\bm{k} 
\sigma}\sum_{\alpha=\pm}\Omega({\bm k})(\gamma\dag_{\bm{k}\sigma\alpha}
\gamma_{\bm{k}\sigma\alpha}+1/2),
\end{equation}
where $\alpha=+$ ($-$) indicates that the momentum is around $+{\bm Q}/2$ ($-{\bm Q}/2$). 
When the gap $\Delta(T)=\om({\bm Q}/2)$ vanishes, $\Omega$'s are the linear dispersions of the Goldstone modes.

{\it Collinear phase}---
To study the instability toward the magnetic ordering, we consider the mean-field decoupling of the interchain/interplane interaction corresponding to the density wave formation of the SB and treat the resulting one/two-dimensional problem \cite{Scalapino, Schulz}. 
The total hamiltonian is given by $H_{1D/2D}=H^{\rm MF}_{1D/2D}+H^{\rm int}$ with
\begin{equation}
H^{\rm int}=
zJ_{\perp} \{|\bm{a}+i\bm{b}|^2-[(\bm{a}-i\bm{b})\cdot \bm{S_Q}+{\rm h.c.}]\},
\end{equation}
where $z$ is the coordination number along interchain/interplane direction and 
$\bm{S}_{\bm Q}=\sum_{k}b\dag_{\bm{k+Q},\sigma}(\bm{\sigma}_{\sigma\sigma'}/2)b_{\bm{k}
\sigma'}$. 
Here $\bm{a}$ and $\bm{b}$ are mean fields for $\la S_{\bm Q}\ra={\bm a}+i{\bm b}$, and collinear and helical orders are expressed 
by them as $\la \bm{a}\times\bm{b}\ra=0$ and $\la \bm{a}\times \bm{b}\ra \ne 0$, respectively \cite{Sonoda}. 
The interaction between $\gamma$-bosons, when translated from that between $b$-bosons by Eq.(\ref{bogol}), is enhanced near the bottom of the dispersion, inversely proportional to the gap $\Delta(T)$ in Fig.3.a, inevitably leading to the density wave instability before the occurence of BEC. 
From the rotational symmetry in spin space, we can set $a^z=b^z=0$ without loss of generality. 
By introducing $s=(a^x-ia^y)+i(b^x-ib^y)$ and $t=(a^x-ia^y)-i(b^x-ib^y)$, we can rewrite $H^{\rm int}$ as
$
H^{\rm int} \sim (zJ_{\perp}/2)\sum_{\bm{k}\sim 0} \{ |s|^2 -
(s b_{{\bm k}-{\bm Q}/2\ua}\dag b_{{\bm k}+{\bm Q}/2\da}+{\rm h.c.}) \}
+(zJ_{\perp}/2)\sum_{{\bm k}\sim 0} \{|t|^2 -(t b_{{\bm k}+{\bm Q}/2 \ua}\dag b_{{\bm k}-{\bm Q}/2 \da} +{\rm h.c.}) \}. 
$
The summations over $\bm{k}$ are restricted to around $0$ since our continuum model is valid only in the low-energy region.  
The free-energy density corresponding to the Hamiltonian $H=H^{\rm MF}_{1D/2D}+H^{\rm int}$ can be written in a decoupled form: $f(x^2)+f(y^2)$, where $x=zJ_{\perp}|s|$ and $y=zJ_{\perp}|t|$.  
Since the helical order is related to $x$ and $y$ through $\la \bm{a} \times \bm{b} 
\ra \propto x^2-y^2$, we conclude that the collinear phase appears if $f(x^2)$ has a global minimum at $x^2 \ne 0$. 
In the quasi-1D case, $f(x^2)-f(0)$ can be expanded in terms of $x^2$ as $Ax^2+Bx^4$ with
\begin{eqnarray}
A&=&\frac{1}{\Delta(T)} \Big( \frac{\Delta(T)}{2zJ_{\perp}}-\frac{S+1/2}{8\delta(T)} \Big), \nonumber \\
B&=&\frac{1}{\Delta(T)^3} \Big( S+\frac{1}{2}\Big) \frac{\delta(T)^3}{128} \Big( 9\frac{(1-2\delta(T)^2/3)^2}{1-\delta(T)^2}-5 \Big), \nonumber
\end{eqnarray}
where $\delta(T)=\Delta(T)/\tilde{\lambda}(T)$ (${\tilde \lambda}(T)=\lambda-2\eta_1 \cos(q/2)$) is the renormalized gap. Here we have assumed $T \gg \Delta(T)$. Since $B$ is positive for $0<\delta(T)<1$, a sufficient condition for the collinear phase is  $A<0$ and a second order phase transition to the collinear state  occurs at $A=0$.  
Above $T_{\rm RVB}$, $\delta(T) \sim 1$ and hence the inequality $A<0$ is not satisfied for small $zJ_{\perp}$. This means $T_{\rm N}<T_{\rm BEC}$, where $T_{\rm N}$ is the antiferromagnetic transition temperature.
Further lowering the temperature with increasing $x$, the gap collapses to result in BEC of SB. 
Therefore, we conclude $T_{\rm BEC}<T_{\rm N}<T_{\rm RVB}$.  
We have also checked the existence of the collinear phase for quasi-2D case by numerically solving the self-consistent equations {\it without} using the continuum model. 
In this way, the instability towards the collinear order is a robust feature of the strongly fluctuating quantum helimagnets, and is essentially different from that of classical system with an easy axis anisotropy.
Now we describe the collinear state $\bm{a}=\bm{b}=(0,a^y,0)$ (see Fig.\ref{q1d2d_lattice}). 
where the 4-fold degeneracy for the energy of 
$\gamma_{\bm{k}\sigma\alpha}$ is split into upper and lower branch bands (see 
Fig.\ref{disp}.b). 
The lower-branch band consisting of linear combinations of 
$\gamma_{\bm{k}\sigma\alpha}$ is 2-fold degenerate. 
The lower branch bosons, $\alpha_{\bm k}$ and $\beta_{\bm k}$, are defined through 
the Bogoliubov transformation as
$
\alpha_k=\cosh\varphi_k (\gamma_{k \ua +}+\zeta \gamma_{-k\da -})/\sqrt{2}-\sinh\varphi_k (-\gamma\dag_{k\da -}+\zeta \gamma\dag_{-k\ua +})/\sqrt{2},
\beta_k=\cosh\varphi_k (\gamma_{k \ua -}+\zeta^* \gamma_{-k \da +})/\sqrt{2}-\sinh\varphi_k (\gamma\dag_{k \da +}+\zeta^* \gamma\dag_{-k \ua -})/\sqrt{2},
$
where $\zeta=e^{i\pi/4}$
and $\tanh 2\varphi_k =x\tilde{\lambda}(T)/(2\Omega(k)^2-x\tilde{\lambda}(T))$
Below, we will focus on the low energy dynamics, and neglect the upper-branch bosons. This leads to the relation between the original bosons 
$b_{\bm{k}\sigma}$: $b_{-\bm{Q}/2+\bm{k}\ua}\sim \zeta^* b_{\bm{Q}/2+k\da}$ and 
$b\dag_{-\bm{Q}/2+\bm{k}\da}\sim -\zeta b\dag_{\bm{Q}/2+k\ua}$.

{\it Helical phase}---
Next we consider the BEC of the lowest modes $\alpha_0$ and $\beta_0$. This 
corresponds to the non-zero expectation values of $b_{\pm{\bm Q}/2,\sigma}$ ($|\la b_{{\bm Q}/2, \sigma} \ra|=|\la b_{-{\bm Q}/2,-\sigma} \ra|$). 
We obtain the cycloidal helical spin structure as
\begin{eqnarray}
S^b_i &\sim & -\sin(\bm{Q}\cdot\bm{R}_i+\pi/4) (|\la b_{\bm{Q}/2 \ua}\ra|^2-|\la b_{\bm{Q}/2\da}\ra|^2)/\cal{N}, \nonumber \\
S^c_i &\sim & S \cos(\bm{Q}\cdot\bm{R}_i+\pi/4), \nonumber \\
S^a_i &\sim & \sin(\bm{Q}\cdot\bm{R}_i+\pi/4) (\la b_{\bm{Q}/2 \ua}\ra \la b_{\bm{Q}/2 \da} \ra+{\rm c.c.})/\cal{N}.
\end{eqnarray}
Here we have used the relaxed constraint $\sum_{i \sigma} b_{i\sigma}\dag b_{i\sigma}=2S {\cal N}$. 
Now we clarify the relation between the elliptic ratio and the Bose condensate fraction. 
If we assume that $\la b_{\bm{Q}/2 \da}\ra=0$ while $\la b_{\bm{Q}/2 \ua}\ra \ne 0$, $S^a_{i}$ becomes zero and the elliptic ratio is given by $m_b/m_c \sim |\la b_{\bm{Q}/2\ua}\ra|^2/({\cal N}S)$. In this case, the spins are rotating counterclockwise within the $bc$-plane. 
The clockwise helicity is realized when $\la b_{\bm{Q}/2 \ua}\ra=0$ while $\la b_{\bm{Q}/2 \da}\ra \ne 0$.
Note that the elliptic ratio can be much smaller than unity even at zero temperature due to the strong quantum fluctuation in sharp contrast to the classical case. 

{\it Neutron scattering spectra}---
Now we turn to the neutron scattering spectra in the helicall phase. For simplicity, we shall focus on the quasi-1D case with the possible relevance to the recent experiment on LiCu$_2$O$_2$ \cite{Seki}. 
The magnetic cross section for polarized neutron is given by the following correlation functions as
$(\frac{d\sigma}{d \Omega} )_{\pm}(q) \propto \la S^{\pm}_{q} S^{\mp}_{-q} \ra$, where the sign $+$ ($-$) corresponds to the parallel (anti-parallel) neutron spin ${\bm S}_n$ to the $a$-axis (see Fig. \ref{q1d2d_lattice}). 
To break the degeneracy of the helicity, we first set $\la b_{\bm{Q}/2 \ua}\ra \ne 0$ and $\la b_{\bm{Q}/2 \da}\ra=0$, i.e., counterclockwise one. We should note here that non-Bragg part is considered below, i.e., non-zero $q$ component.
In the low energy regime, using $\alpha_{\bm k}$ and $\beta_{\bm k}$, 
$\la S^{+}_{Q+q} S^{-}_{-Q-q} \ra \sim F_1/q^2 + F_2/q$ and $\la S^{+}_{-Q-q}S^{-}_{Q+q} \ra \sim F_2/q$, respectively, with
\begin{equation}
F_1=\frac{|\la b_{{\bm Q}/2 \ua} \ra|^2}{\cal N} \Big(S+\frac{1}{2} \Big)\frac{\Delta(T)}{2c_{\parallel}}, \:
F_2=\Big( S+\frac{1}{2} \Big)^2 \frac{\Delta(T)^2}{16c^2_{\perp}}, \nonumber
\end{equation}
where $q$ is assumed to be small. The difference $\la S^{+}_{Q+q} S^{-}_{-Q-q} \ra-\la S^{+}_{-Q-q}S^{-}_{Q+q} \ra$ is expressed by the vector spin chirality $\la ({\bm S}_{Q+q}\times {\bm S}_{-Q-q})^z \ra/i$, and is directly related to the condensate fraction, i.e., the $F_1$ term. 
The crossover between the $F_1$ and $F_2$ terms occurs at $q_c a \propto |\la b_{{\bm Q}/2 \ua} \ra|^2 \frac{c_{\perp}^2/a}{c_{\parallel} \Delta(T)} \sim (\frac{J_{\perp}}{J_{\parallel}})^3 \frac{J_{\parallel}}{\Delta(T)}$, where $a$ is the lattice constant and $J_{\parallel}$ is the typical energy scale determined by $J_1$ and $J_2$. 
Another important correlation functions, $\la S^{\alpha}_q S^{\alpha}_{-q} \ra$ ($\alpha=x,y$), can be observed by the ${\bm S}_n \parallel c$ setup. 
By a similar calculation, one can show that $\la S^x_{Q+q} S^x_{-Q-q} \ra = \la S^y_{Q+q} S^y_{-Q-q} \ra$ for the fluctuating part. 
In the experiment \cite{Seki}, $\la S^{\pm}_{Q+q} S^{\mp}_{-Q-q}\rangle$ suggests elliptic spiral while $\la S^{\alpha}_{Q+q} S^{\alpha}_{-Q-q} \ra$ indicates circular one. This puzzling point 
would be resolved by our above anaysis considering the quasi-elasitic component \cite{Furukawa}. 

\begin{figure}
\includegraphics[width=8.9cm]{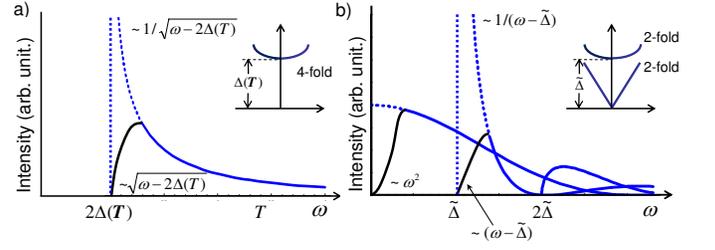}
\caption{Schematic plots of Im$\epsilon_{aa}(\omega)$ in a) the paramagnetic phase and b) the helical phase with $T=0$
($\tilde{\Delta}=\sqrt{2}\Delta(0)$). 
Behaviors nearly thresholds are indicated.
Blue (solid and dotted) lines are the results for purely 1D model.
Singularities are smeared out by the interchain interaction as shown by black lines. 
Insets are schematic boson dispersions in both the phases.}
\label{Im_eps}
\end{figure}

{\it Dielectric response}---Finally, we examine the dynamical dielectric response both in the paramagnetic and helical phases of the quasi-1D model. 
Even in the paramagnetic and collinear phase, we assume that 
the fluctuating electric polarization is given by Eq. (\ref{KNB}) \cite{Miyahara}. 
We take the mean-field decoupling  $\bm{S}_i\times\bm{S}_j 
= (\la b_{i\mu}\dag b_{j\mu}\ra (b_{j\rho}\dag\bm{\sigma}_{\rho\nu}b_{i\nu})-{\rm 
h.c.})/(4i)$ to the ferromagnetic bonds and $\bm{S}_i\times\bm{S}_j =(\la b_{i\mu}\dag\epsilon_{\mu\rho}b_{j\rho}\dag \ra (b_{j\sigma}\bm{\sigma}^*_{\sigma\lambda}\epsilon_{\lambda\nu}b_{i\nu})-{\rm h.c.})/(4i)$ to the antiferromagnetic bonds.  
We henceforth focus on the contribution from the antiferromagnetic ($J_2$) bonds since its fluctuation is stronger than that of the ferromagnetic one. 
From the geometry of the system (see Fig.\ref{q1d2d_lattice}), the polarization along the $b$-axis $P^b$ is always zero. 
In the paramagnetic phase, Im$\varepsilon_{aa}(\omega)=$Im$\varepsilon_{cc}(\omega)$ due to the rotational symmetry in spin space. 
The expression for the polarization along the $a$-axis $P^a$ is given by
\begin{equation}
P^a \propto (\eta_2/J_2)\sum_{\bm k}\cos (2k_x)(ib_{{\bm k}\sigma}b_{-{\bm k}\sigma}+{\rm h.c.}).
\nonumber
\end{equation}
For purely 1D case, ${\rm Im}\varepsilon_{aa}(\omega) \propto(n(\omega/2)+1/2)/(\omega \sqrt{\omega^2-4\Delta(T)^2})$, where $n(\omega)$ is the Bose distribution function. 
Near the threshold of the absorption, the 1D van Hove singularity, ${\rm Im}\varepsilon_{aa}(\omega) \propto 1/\sqrt{\omega-2\Delta(T)}$, appears as schematically shown in Fig.\ref{Im_eps}.a. 
On the other hand, a drastic change of the absorption spectra occurs in the helical phase since the low-lying branch bosons become gapless (see Fig.\ref{disp}.c). 
In this phase, the energy dispersions of the upper and lower branches are given by $\Omega_+(k_x)=\sqrt{c^2_{\parallel}k^2_x+2\Delta(0)^2}$ and $\Omega_-(k_x)=c_{\parallel}k_x$, respectively. 
We assume the BEC of SB by the weak interchain interaction. 
The schematic behavior at zero temperature is shown in Fig.\ref{Im_eps}.b. 
There are three contributions corresponding to the processes of two bosons 
i) in the upper branch, ii) in the gapless-lower branch and iii) in both the upper and lower branches, respectively. 
Finally, it should be noted that we cannot neglect the one-magnon contribution coming from the condensed portion in the helical phase. 
This contribution corresponds to that obtained in the previous analysis \cite{colmod}, but this is  much smaller in the quantum limit. 

The authors are grateful to 
S. Seki, Y. Yamasaki, N. Kida, S. Todo, and Y. Tokura for fruitful discussions.
This work was supported in part by Grant-in-Aids (Grant No. 15104006, No. 16076205, and No. 17105002) and NAREGI Nanoscience Project from the Ministry of Education, Culture, Sports, Science, and Technology.
H.K. and S.O. were supported by the Japan Society for the Promotion of Science. 


\begin{thebibliography}{99}
\bibitem{Mochizuki}
Although we can also apply this mean-field method to the 3D systems, the results are almost the same as those obtained by classical $+$ spin wave theory.
\bibitem{collapse}
With moderate 3D couplings, our SB mean-field theory can give the helical magnetism even in quasi-1D half-integer spin systems.
\end{thebibliography}
\end{document}